\documentclass[a4paper,11pt]{article}
\usepackage{aaskaiid}
\usepackage{aas-macros}
\usepackage{orcidlink}

\newcommand\degr{\ensuremath{^\circ}}
\newcommand\arcmin{\ensuremath{^\prime}}
\newcommand\arcsec{\ensuremath{^{\prime\prime}}}

\newcommand{\fig}{Fig.}

\newcommand{\sect}{Section}
\newcommand{\sects}{Sections}

\newcommand{\eqn}{Equation}

\newcommand{\Figs}{Figs.}

\title{Methods of Observing and Characterising the Ionosphere with SKA-Low}
\ShortTitle{Characterising the ionosphere}

\author[1]{John~S.~Morgan\,\orcidlink{0000-0001-9224-5483}}
\ShortName{Morgan~et~al.} 
\author[2]{Biagio~Forte\,\orcidlink{0000-0003-1682-1930}}
\author[3]{Kshitija~B.~Deshpande\,\orcidlink{0000-0001-5987-9142}}
\author[4]{Andrzej~Krankowski\,\orcidlink{0000-0003-2812-6222}}
\author[5]{Mario~M.~Bisi\,\orcidlink{0000-0001-6821-9576}}

\affiliation[1]{CSIRO, Space and Astronomy, P.O. Box 1130, Bentley, WA 6102}
\emailAdd{john.morgan@csiro.au}
\affiliation[2]{Department of Electronic and Electrical Engineering, University of Bath, BA2 7AY, Bath, UK}
\emailAdd{B.Forte@bath.ac.uk}
\affiliation[3]{Department of Physical Sciences, Embry-Riddle Aeronautical University, Daytona Beach, FL 32114, USA}
\affiliation[4]{Space Radio-Diagnostics Research Centre, University of Warmia and Mazury in Olsztyn, Poland}
\affiliation[5]{RAL Space, United Kingdom Research and Innovation – Science \& Technology Facilities Council – Rutherford Appleton Laboratory, Harwell Campus, Oxfordshire, OX11 0QX, UK}

\abstract{
The ionosphere and its behaviour critically affects ground-based radio instruments at low frequencies, and radio interferometry has been used as a probe of the ionosphere since the earliest days of radio astronomy.
In this chapter, we aim to give an overview of the ionosphere and its salient properties in the mid-latitudes where the SKAO instruments are located.
We provide a comprehensive review of its impact on the astrophysical radio signals which traverse it.

We then focus on the ionosphere as a phase screen, and the many ways in which the ionospheric structure can be measured using a low-frequency interferometer such as SKA-Low.
Our aim here is to provide the broadest possible spectrum of measurement approaches.
We place particular emphasis on the wide range of innovative approaches that have been developed for SKA precursors and pathfinders over the last decade, however we also draw attention to other approaches, some untested, that appear in the literature.

Next, we consider an innovative approach for deducing the detailed physical conditions in the ionosphere from SKA observables via iterative simulations with a sophisticated physical model from which the interferometric response can be forward-modelled. This approach has proven extremely successful for interpreting large scale observations in the complex polar region of the ionosphere, and we discuss how it can be applied to the SKA-Low.

Finally, we provide a summary of the technical requirements which will ensure viability of the various techniques discussed.
}

\begin{document}
\maketitle

\section{Introduction}
\label{sec:intro}
Low-frequency radio astronomy and ionospheric research have been intertwined since the very earliest radio telescopes.
Observations of the ionosphere in the 1930s and onwards often studied the interference between ground waves and ionosphere-reflected radio waves \citep{1933PCPS...29..301R}.
This ``Lloyd's Mirror''-inspired approach \citep{Lloyd1831} then spurred the new science of Radio Astronomy \citep[][p.~193]{2023bea..book...60G}, which was soon being used to make novel measurements of the ionosphere \citep{1948TeMAE..53..429P}.

There are many reasons for characterising the ionosphere, aside from studying it for its own sake \citep{materassi2019dynamical,ishii2024space}.
Ionospheric effects continue to pose a challenge to the operation of key infrastructure, and forecasting and monitoring remain constrained by lack of data \citep{Tsagouri2023}.
Another motivation is mitigating the impact of the ionosphere on the SKA telescopes themselves, ensuring that they reach their full potential.
Detailed studies may provide insights which inform calibration and imaging strategies \citep[e.g.][]{2010ISPM...27...30W,2005ASPC..345..317E}; real-time monitoring of the ionosphere above the SKA-Low telescope would facilitate dynamic scheduling\footnote{AKA ``smart scheduling'', ``flexible scheduling''} to optimise science outcomes, as is routinely done for shorter wavelengths on the basis of prevailing tropospheric weather conditions \citep[e.g.][]{2009ASPC..411..330B} or seeing \citep[e.g.][]{2018MNRAS.480.1278O}.

The ionosphere imposes a variety of effects on trans-ionospheric radio waves from astrophysical sources.
In this chapter we will focus on the effects induced by the ionosphere as a phase screen.
Immediately after traversing a medium of varying refractive index, a planar wave (such as one from any astrophysical source) will have a uniform amplitude, but with phase distortions exactly matching those imparted by the screen.
However, at increasing distance from the phase screen, the phase fluctuations will develop through interference into amplitude fluctuations, known as amplitude scintillation \citep{Uscinski1977}.
The resulting diffraction appearing in a pattern on the ground can then be measured in order to recover the structure of the ionosphere\footnote{Spaced receivers can also be used to measure the diffraction pattern of terrestrial signals reflected from the ionosphere \citep{1950PPSB...63..106B}, which motivated the construction in Australia of a square kilometre array operating at 2\,MHz \citep{1969Natur.223.1321B}. The SKA-Low, operating at a frequency 2 orders of magnitude higher, will be able to probe much finer structures.}.

This chapter is organised as follows:
in \sect~\ref{sec:ionosphere} we describe the ionosphere and its effects on electromagnetic radiation;
in \sect~\ref{sec:methods} we explain the wide variety of methods by which the ionosphere can be observed using radio interferometers;
in \sect~\ref{sec:simulations} we explain how interferometric observables can be linked to sophisticated models of the ionosphere;
and in ~\sect~\ref{sec:roadmap} we provide an overall approach to observing the ionosphere with SKA-Low, and note the technical requirements for each observing approach.

Phase effects in a dispersive medium scale with $\lambda^2$, and scintillation scales with $\sim$$\lambda$, so ionospheric effects are much more profound in the SKA-Low observing band.
Therefore, although SKA-Mid and SKA-Low will both be affected by the ionosphere, in this chapter we restrict our discussion to SKA-Low.
However, the effects (and some of the observing techniques) discussed here will also be applicable to SKA-Mid, especially (though not exclusively) below 1\,GHz.

There are two other topics that are highly pertinent to the subject of this chapter, that we have had to exclude from detailed discussion in order to keep the the chapter to a reasonable length. 
These are Faraday rotation induced by the ionosphere, and synergies of SKA-Low observations with other instruments used for probing the ionosphere.

\section{General properties of the Ionosphere}
\label{sec:ionosphere}
\subsection{Ionospheric structure}
The ionosphere is a part of the Earth’s atmosphere where neutral species are ionised by solar radiation at various wavelengths. This ionisation gives rise to plasma encircling Earth, which is subject to the action of the geomagnetic field, electric fields, and drifts in neutral species. The ionospheric plasma extends approximately between 80 km to 1000 km of altitude, and it is typically subdivided in layers (D, E, F, topside) in view of different processes prevalent at various altitudes. For example, collisions with neutral species and recombination prevail at lower altitudes whereas diffusive equilibrium is typical of higher altitudes. The plasma density generally increases with altitude and maximises in the F region before a general decrease in the topside where the ionospheric plasma converges into the outer parts of the atmosphere (e.g., the magnetosphere) \citep{kelley2009earth}.

The balance between ionisation, recombination, and diffusion leaves a magnetised plasma that responds to the influence of the combined action of geomagnetic field, electric fields, neutral winds, and varying space weather conditions. The distribution of plasma density in the ionosphere varies in space and evolves in time. The smooth spatial distribution of the bulk ionisation can be accompanied by inhomogeneities that form as a consequence of plasma instability mechanisms. The magnetohydrodynamics of these mechanisms originate inhomogeneities (or irregularities) that can also evolve in space and time \citep{kelley2009earth}.

At equatorial latitudes\footnote{SKA-Low and SKA-mid are by design located at mid-latitudes where the ionosphere is \emph{relatively} benign. SKA-low has a magnetic latitude of -60, SKA-Mid has a magnetic latitude of -64.} irregularities mainly form in longitudes corresponding to post-sunset and nighttime, where the ionosphere can be lifted towards higher altitudes across the equatorial anomaly by a change in direction of the electric field. The uplifting of dense plasma generates instabilities typically distributing around plasma bubbles (structures of depleted plasma extending to higher altitudes and across the equatorial anomaly crests). At auroral and polar latitudes irregularities mainly form in conjunction with plasma patches that form from the tongue of ionisation (a region of enhanced plasma density extending from the dayside into the nightside typical of disturbed magnetic conditions) and drift across the polar cap and into the auroral oval following various convection patterns. Irregularities also form as a consequence of particle precipitation which varies with the energy of precipitating charged particles. 

With the advent of adverse space weather conditions (and depending on magnetic reconnection on the dayside) the high latitude ionosphere is exposed to increased ionisation (which starts on the dayside and protrudes towards the nightside, causing plasma patches to form and drift across the polar cap) and to increased particle precipitation (which can induce additional ionisation at various altitudes depending on the energy of precipitating particles). As a consequence of modified magnetic conditions, the enhancement in energetic particle precipitation can distribute over a wider range of latitudes, hence causing the auroral oval to expand equatorward (with particle precipitation occurring at more equatorward latitudes than under quiet conditions). Particle precipitation induces optical emissions and irregularities in the plasma density distribution. The amount of equatorward expansion (and corresponding optical emissions, enhancement in ionisation, and altitudes over which these happen) depends on the magnitude of events of solar origin such as coronal mass ejections (CMEs) and how they modify the interplanetary magnetic field and its reconnection with the geomagnetic field.

On the other hand, at equatorial latitudes the uplifting of the ionosphere can be inhibited, hence causing a suppression of the onset of irregularities. 

Although the macro-scale morphology of the ionospheric response is well understood on average, the precise way irregularities distribute in space and evolve in time during magnetic storms and substorms varies on a case-by-case basis, and it represents an outstanding challenge. 

\subsection{Overview of ionospheric impacts}
Ground-based radio astronomy is possible because radio waves are able to traverse the Earth's atmosphere largely unimpeded.
The bottom edge of this ``radio window'' \citep[e.g.][]{2016era..book.....C} is set by the ionosphere, which, at frequencies below the bottom edge of the SKA band, can completely absorb, or reflect (internally refract) electromagnetic radiation.
These effects (and ionospheric emission) become important close to the plasma frequency of the ionosphere, whose peak (FoF2) is below 10\,MHz \citep[e.g.][]{10.1049/PBEW031E}.

Radio waves propagating through the ionosphere experience a refractive index that is a function of the combination of features related to the magnetic field and to collisions. Radio waves propagating through the ionosphere can be originated from artificial satellites (such as in the case of satellite telecommunications and satellite navigation) or from astrophysical sources (such as those utilised in radio astronomy). The ionosphere is dispersive for radio waves as the refractive index also varies with the radio wave frequency. The radio wave frequencies that experience most ionospheric propagation effects are those between HF and C band: at higher frequencies ionospheric propagation effects reduce with frequency until they become negligible. The full description of the refractive index in the ionosphere is provided by the Appleton-Hartree equation \citep{appleton:1932}. The general result of the Appleton-Hartree equation can be approximated when some features can be neglected, hence allowing the description of propagation effects in a simpler fashion without loss of generality.

In general, radio waves at relevant radio wave frequencies propagating through the ionosphere experience an advance in phase and a group delay. The magnitude of these two effects is the same but with opposite sign. The magnitude is directly proportional to the electron density integrated along the ray path in the ionosphere: the Total Electron Content (TEC). 

The TEC is a useful quantity that can be extracted from satellite radio signals and that can be utilised to provide a measure of the ionosphere along a particular direction. Over recent years, the use of satellite navigation signals (from Global Navigation Satellite Systems – GNSS) has allowed estimation of TEC along various directions at the same time (from a ground station receiving radio signals from several GNSS satellites at the same time): this information can be combined to provide an estimate of the ionospheric TEC along the zenith. The combination of this information from multiple ground stations distributed across the planet allows for a global estimate of the TEC, hence providing a way to monitor the state of the ionosphere over the entire planet and how it evolves in time.

Interferometers do not usually sense absolute phase, but are exquisitely designed to measure the phase difference between individual elements.
To first order, these phase changes consist of a ``wedge''; i.e. a phase gradient across the array.
This corresponds precisely to a refractive shift in position in the image plane.
Just as with optical wavelengths, sources observed close to the horizon will experience a large refractive shift due to the curvature of the Earth\footnote{This is quantified by \citet[][see equations 21\&29, and \fig~5]{1960AuJPh..13..153K}. At 50\,MHz, assuming a maximum FoF2 of 8\,MHz, the effect is $\sim$2\arcmin\ at 30\degr\ elevation. Note that the ionosphere, being a dispersive medium, induces refraction \emph{towards} the direction of lower electron density \citep[e.g.][\eqn~1]{2015RaSc...50..574L}. This effect may be further reduced by standard self-calibration techniques unless a calibrator nearer the zenith is used for a target much closer to the horizon or vice versa.}.
Unlike at optical wavelengths, diurnal variations \citep{1952JATP....2..350S} as well as other large-scale features such as Travelling Ionospheric Disturbances (TIDs) can also cause large phase gradients which shift the positions of sources very noticeably.

Just as interferometers are designed to measure phase differences between elements, polarimetrically capable instruments can also sense orthogonal polarisations separately, and the phase relationship between them.
Thus, for interferometers observing linearly polarised sources, an important ionospheric propagation effect is that of Faraday Rotation (FR). The electric field of polarised radio waves traversing the ionospheric magnetised plasma rotates by an amount proportional to the integral of the product between the electron density and the component of the magnetic field along the ray path (a higher order effect arising from the solutions of the Appleton-Hartree equation). In the early days of ionospheric radio science enabled by the first artificial satellites, this effect was utilised to estimate TEC under some assumptions. FR decreases with radio wave frequency. The use of circular polarisation in satellite radio signals has made this effect negligible. However, FR represents a powerful effect that is utilised in radio astronomy, for example, to infer information about the magnetic field in galaxies. In this case, the FR is determined by the plasma distribution and magnetic field in these astrophysical regions. For observations carried out through radio telescopes on Earth, the ionosphere can still contribute to the overall FR, although the magnitude of the ionospheric contribution typically is smaller than the FR from astrophysical regions: however, the ionospheric contribution still needs proper calibration.

If traversing ionospheric irregularities, the wavefront of radio waves is distorted: the changes in phase along the wavefront build a diffraction pattern with the propagation distance. To an observer on Earth this diffraction pattern can manifest as a temporal pattern if there is a relative drift between the ray path and the ionospheric irregularities: this effect is known as radio wave scintillation \citep{forte2022interpretation}. In general, scintillation decreases with increasing radio wave frequency. At L band, scintillation can be more intense at high latitudes and at equatorial latitudes (where it can still be strong) \citep{forte2012analysis}. At lower radio wave frequencies, scintillation can also manifest at middle latitudes more often. In the presence of scintillation both the phase and the intensity of the received radio waves fluctuate in time \citep{forte2008refractive}. 

In the case of radio waves from artificial satellites and measured by means of demodulation in radio receivers, intensity scintillation induces fading that can lead to the complete loss of a signal in the most intense cases (typical at equatorial latitudes). 

Radio wave scintillation can be observed and measured on radio waves emitted from astrophysical sources and measured by means of radio telescopes \citep{flisek2023towards,forte2022interpretation}. This is primarily observed at frequencies between the VHF and L band. Scintillation on radio waves from astrophysical sources has been observed by utilising beam-formed observations of specific objects. However, inteferometric imaging (or multi-channel beamforming) approaches allow multiple objects to be observed simultaneously, and since the sky is filled with discrete radio sources, this is potentially a very powerful technique for measuring the ionosphere on multiple sightlines, limited only by the sensitivity and field of view of the instrument.

Radio wave scintillation can be utilised to infer information about irregularities forming in the ionosphere and to understand the mechanisms regulating their evolution in space and time. Contrary to satellite observations that are available at few and specific frequencies, modern interferometers offer the possibility of measuring scintillation over a much wider range of radio wave frequencies, hence enabling the detection of irregularities with gradients occurring over multiple spatial scales. This capability is instrumental for understanding factors determining the cascade (or its absence) from larger to smaller spatial scales as irregularities evolve in space and time in the ionosphere. This experimental information can guide models to describe processes in the ionosphere in a more comprehensive fashion and to address the outstanding challenge of ionospheric prediction together with the prediction of the ionospheric impact in the presence of varying space weather conditions.

\subsection{The ionospheric diffraction pattern}
\label{subsec:diffraction}
Scintillation can be described as originated by a form of scattering. This scattering is the change in phase introduced by irregularities along the ray path, and it reflects the fact that irregularities tend to adopt a patchy and stochastic distribution in space. The amount of scintillation arising from the propagation through ionospheric irregularities can, in principle, be low or high (this is regularly observed by means of satellite radio receivers). 

The case of low scintillation can be described by approximating that all the phase changes along the wavefront occur in a very limited region of the propagation medium. This approximation is known as a phase-changing screen where the scattering is weak (and localised to the phase screen only). This is a mathematical approximation that allows the description of intensity and phase fluctuations in a statistical sense through the Rytov solution. Although a very good approximation, caution must be exerted when considering this description, which applies only in the case of very small phase and intensity fluctuations, with weak scattering.

When scintillation increases in level, this approximation is no longer valid and scattering occurs in many points along the ray paths. In this case, the mathematical discussion needs to account for multiple scattering that originates from irregularities distributed along an extended interval of the ray path. 

By measuring scintillation and its spectral properties it is possible to deduce properties of the spatial distribution of irregularities in the ionosphere. Modern and future interferometers offer the capability to compare simultaneous observations over a wide range of radio wave frequencies from closely located stations, distributed over various distances (or baselines) with varying directions and magnitudes. This fact can be exploited to reconstruct the statistical properties of the ionosphere to understand how irregularities distribute in space and evolve in time. In the case of the SKAO, the number of stations together with the distribution and extent of the baselines are much larger than any existing pathfinder, hence offering a much higher resolution in reconstructing the distribution of ionospheric irregularities. Contrary to traditional satellite observations which are convoluted with the motion of satellites, modern interferometers can truly sound the ionosphere over multiple spatial scales and in time\footnote{The apparent motion of astrophysical sources (due to Earth Rotation) is non-negligible, although much slower than that of emitters in low Earth orbit, being approximately 20\,m\,s$^{-1}$ at a height of 300\,km for a source at the zenith above SKA-Low}. Their limitations are represented by the fact that the radio wave frequencies are within a certain range and do not cover the full HF-to-C band interval at the same time and in the same way.

Two quantities are useful in describing the strength of scattering.
The variance of the phase screen can be quantified by the diffractive length scale, $r_\mathrm{diff}$.
This is the transverse separation over which the root mean square difference in phase reaches 1\,radian (smaller numbers equate to a stronger scattering screen).
As outlined above, the diffraction pattern builds up as the wave propagates away from the phase screen.
This geometric effect is captured by the Fresnel scale, $r_F$, which is proportional to the square root of both the observing wavelength, and the distance to the scattering screen.
A useful measure of the amplitude scintillation strength is then given by $r_F/r_\mathrm{diff}$.
The reader is referred to \citet{1992RSPTA.341..151N} for a simple but comprehensive review of amplitude scintillation for a variety of astronomy-related cases, including such concepts as the Fresnel scale, Kolmogorov turbulence, scintillation regimes, and the effect of source size (see also \sect~\ref{sec:simulations} in this chapter).

An important point of difference between the phase and amplitude of the diffraction pattern on the ground is the so-called ``Fresnel Filter''.
The 2D power spectrum of the amplitude fluctuations as a function of wavenumber $q$ is modulated by a $\sin^2q^2$ function, whereas that for the phase fluctuations is modulated by a $\cos^2q^2$.
This is discussed further in \sect~\ref{sec:simulations}.

Another important consideration when treating scintillation observed on radio waves emitted from astrophysical sources is the angular extension of the source. This represents a key difference from traditional ionospheric satellite observations (where the transmitter can be approximated as a point source). The effect of the source extension needs to be taken into account to disentangle effects related to the source from effects related to the propagation through ionospheric irregularities.

\subsection{Advantages of observing with a radio interferometer}
\citet{2015RaSc...50..574L} provide an overview of ionospheric observables, both using radio telescopes and other instruments.
Here we focus on the increased sensitivity offered by low-frequency radio interferometry.

Scintillation effects induced by the ionosphere are regularly observed using Global Navigation Satellite Signals (GNSS). 
However, the gigahertz frequencies utilised mean that these measurements are typically insensitive to the more benign scintillation effects which prevail at the mid latitudes. 
Radio beacons can provide some insights \citep[][and references therein]{2023AdSpR..72.5503L,forte2008refractive} but constellations that compete with the number of astrophysical sources detectable by a sensitive radio telescope have not been built, and would not be able to work at frequencies reserved for radio astronomy.
The broadband nature of cosmic radio emission also opens up analysis techniques which would not be possible with artificial beacons.

As noted above, radio interferometers are able to make exquisitely precise differential phase measurements.
This means that they are able to sense extremely small ionospheric gradients.
For instance, MWA observations of refractive shifts hit a noise floor at $\sim$10\arcsec\ at 150\,MHz \citep{2017MNRAS.471.3974J}, and this corresponds to a spatial gradient of 3$\times$10$^{10}\,$electrons~m$^{-3}$ in SI units\footnote{or, equivalently, 3\,mTECu~km$^{-1}$ where one unit of Total Electron Content is a column density of 10$^{16}$ electrons~m$^{-2}$. Note that Rate of TEC is typically given in TECu per minute \citep[e.g.][]{2014RaSc...49..653C}.} \citep[][\eqn~4]{2015MNRAS.453.2731L}.
This exceptional sensitivity to small gradients means that features that are just detectable with careful measurement of GNSS phase observables can be imaged in detail \citep{2016JGRA..121.1569L}.

\section{Methods for probing the Ionosphere with an interferometer}
\label{sec:methods}
\begin{figure}[ht]
    \centering
	\includegraphics[width=0.8\columnwidth]{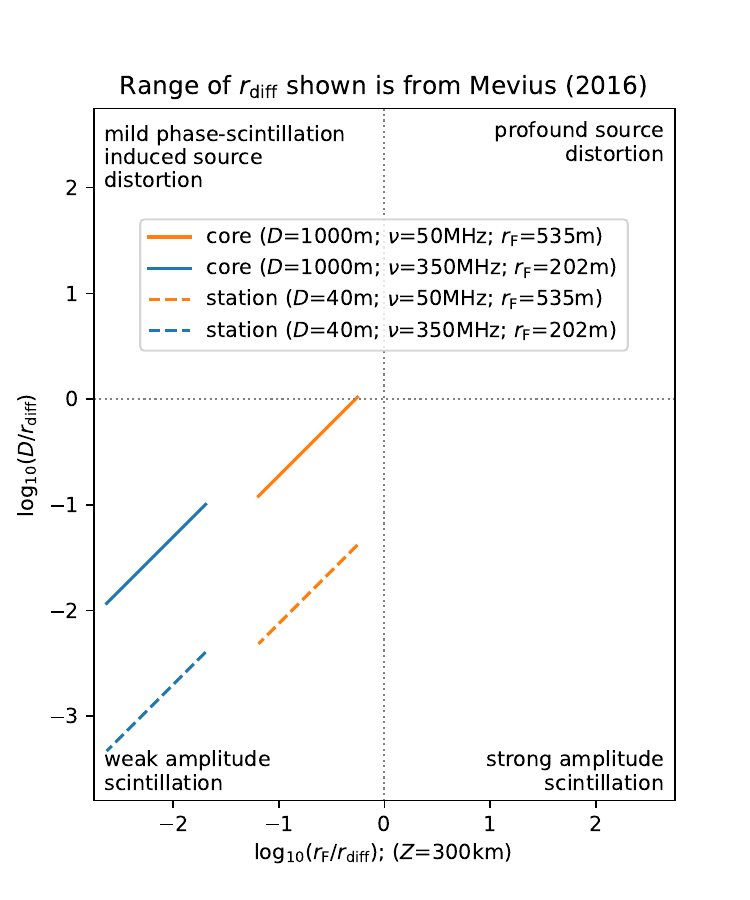}
	\caption{Following \citet{2022PASA...39...36W} and \citet{1989JOSAA...6..977C}. We use the range of $r_\mathrm{diff}$ observed by \citet{2016RaSc...51..927M}. Large $r_\mathrm{diff}$ (weaker fluctuations) are at the bottom-left of each line. Small $r_\mathrm{diff}$ (stronger fluctuations) are towards the top right.}
    \label{fig:cornwell}
\end{figure}

With the picture of the amplitude/phase scintillation pattern on the ground now complete, we can consider the effect this has on an interferometer.
\citet{1989JOSAA...6..977C} provide 4 useful asymptotic regimes for understanding the effect in the image plane.
Note that in contrast to other frameworks, e.g. that of \citep[][see also \sect~2 and \fig~1 of Datta et al. in this volume]{2004:Lonsdale}, this considers both amplitude and phase scintillation effects.
In \fig~\ref{fig:cornwell} we follow \citet{2022PASA...39...36W} by demonstrating how the SKA compares to these regimes for a range of $r_\mathrm{diff}$ observed by \citet{2016RaSc...51..927M}, for the lowest and highest SKA-Low frequency, for both a single station, and across the 1-km diameter core.
This diagram provides a range of valuable insights, which we detail below.

Even at the bottom of the SKA-Low frequency range, in more extreme ionospheric conditions, scatter remains in the weak regime;
although $r_F/r_\mathrm{diff}\gtrsim0.1$ might be better understood as an intermediate regime, and strong scatter has been observed at the SKA-Low's location, albeit rarely \citep[e.g.][]{2015GeoRL..42.3707L,2016JGRA..121.1569L,2025PASJ...77..556Y}.

By design, the 40-m diameter of a single SKA-Low station is sufficiently small that phase changes across it, even in extreme conditions are $\ll$1\,rad.
However, a single station will certainly experience amplitude scintillation, with a modulation index $\sim1$\% at the top of the band, and 10\% or more at the bottom of the band.

In the more active ionospheric conditions, particularly at the lower SKA-Low frequencies, the conditions do not correspond well to any of the 4 asymptotic regimes, and will have features of all four: amplitude scintillation, refractive shifts due to phase gradients across the array, and also focussing/defocussing of sources due to wavefront curvature across the array.

In the remainder of this section we discuss various approaches to sensing the ionosphere with the SKA.

\subsection{Baseline-based phase measurements}
\label{subsec:baseline}
Radio interferometers are outstanding instruments for characterising phase screens such as the ionosphere \citep{2019astro2020T.307B}.
Characteristics which make good imaging instruments (a range of baseline lengths at different orientations) are also what is needed to measure the phase ``structure function'' \citep[see, e.g.][\sect~13.1.16 for a definition and discussion]{2017isra.book.....T}.
This approach has been taken by \citet{2016RaSc...51..927M}, who used Lofar in order to probe the ionospheric phase structure function over three orders of magnitude of baseline from 100\,m to 100\,km.
They found that the ionospheric phase structure function is well-described by power law turbulence, with a power law index slightly steeper than that for Kolmorogorov.
SKA-Low will have baselines which span a similar range, albeit with a larger number of baselines in total.

Astrometric and geodetic Very Long Baseline Interferometry (VLBI) observations must also compensate for ionospheric effects, and so the ionospheric phase structure function has also been measured on VLBI baselines.
These VLBI observables have also been compared with Global maps constructed from observations of Global Navigation and Satellite Systems \citep{2003RaSc...38.1069S}.
\citet{2011AJ....142...35P} show that ionospheric delays are covariant over baselines up to 2000\,km.

Such studies lend themselves to the situation where a very bright calibrator source dominates the visibilities over the entire field of view.
However, in principle the visibilities encode the phase for every baseline for every source in the field, and in principle these quantities may be recoverable, even for the case of the MWA which has a very wide field of view \citep{2022JATIS...8a1012R}.

\subsection{Image-based phase measurements (refractive shifts)}
\label{subsec:shift}
The MWA, with its extremely wide field of view, but relatively small aperture plane (approx. 6\,km in its final configuration) lends itself to measurements of the ionosphere in the image plane.
After imaging, the phase gradient across the array, towards each source, will cause the source to appear in a different location.
The measurement of this offset for many sources across the field of view provides a 2-dimensional vector field of the ionospheric gradient across the full field of view ($\sim$100\,km or more for the MWA).
This approach was used by \citet{2015GeoRL..42.3707L} to show the existence of magnetic field-aligned ducts in the magnetosphere.
If desired, \citet{2017MNRAS.471.3974J} have shown that this 2D vector field can be integrated to provide the 2D scalar electron density (Total Electron Content; TEC) across the field of view; however there is an unknown offset (or DC term) since the constant of integration remains unknown.
This constant could be restored via external measurements, such as GNSS \citep{2015PASA...32...29A}, or via Faraday rotation measurements of natural or artificial radio sources \citep{2017PASA...34...40L,2023AdSpR..72.5503L}

In the case of clear coherent features being seen, a very useful extension to this approach is to use parallax measurements to determine the height of these features. 
This has been achieved for the MWA by dividing the array's elements into two interferometers, repeating the imaging with East and West arrays independently, and then measuring the parallax of features in both images \citep{2015GeoRL..42.3707L}.
These techniques have demonstrated that on occasion very extreme phenomena have occurred over the MWA \citep{2025PASJ...77..556Y}.

Furthermore, the calibration and imaging of data taken for other purposes, such as the Epoch of Reionisation studies \citep{2017MNRAS.471.3974J}, or imaging surveys \citep{2020RaSc...5507106H} can provide these detailed ionospheric data as a by-product, and thus provide novel physical insights into the ionosphere.

\subsection{Amplitude scintillation}
\label{subsec:scint}
Amplitude scintillation probes the smallest scales possible with remote sensing.
At least in the weak scatter regime (see sect~\ref{subsec:diffraction}), spatial scales probed are those close to the Fresnel scale, which varies from $\sim200$\,m--$\sim500$\,m across the SKA-Low band, assuming a height of irregularities of 300\,km.
\citet{2022PASA...39...36W} observed ionospheric scintillation using the 100-m diameter core of the Phase-I MWA.
This work underlines that multiple observables (in this case amplitude scintillation indices and refractive shifts) can be extracted from the same observations, and that the combination of these observables can lead to valuable insights.
For example, in this case, $r_\mathrm{diff}$ can be computed for the two different techniques and compared.
In general, the two are correlated, and show a 1:1 relationship within a factor of a few.

\citet{2022PASA...39...36W} go some way to resolving the slight tension between the Kolmogorov turbulence picture seen in baseline-based measurements (\sect~\ref{subsec:baseline}) and an ionosphere dominated by large-scale, coherent features seen in measurements of refractive shift (\sect~\ref{subsec:shift}).
Since, in most cases, $r_\mathrm{diff}$ computed via refractive shifts on a scale of 2\,000\,m could be reconciled with $r_\mathrm{diff}$ computed via scintillation on a scale of 300\,m via a Kolmogorov power-law structure function, even when the former shows large coherent structures, it would appear that in most cases, even the large coherent structures actually have a fractal, turbulent structure.

Both the refractive shifts and the scintillation indices used by \citet{2022PASA...39...36W} were computed as an ensemble of all sources across the field of view.
More detailed analysis would be possible with a direction-dependent approach.

\subsubsection{The secondary spectrum}
\label{subsubsec:2spectrum}
\citet{2014JGRA..11910544F} have probed the ionosphere in greater detail, utilizing the Fourier transform of the dynamic spectrum of scintillations (the so-called ``secondary spectrum'').
In the strong regime (or transition to the strong) scintillation has a lot of structure\footnote{Although it should be possible to make the same measurements in the weak regime, albeit with lower signal-to-noise due to the weaker scattered component \citep{2006ApJ...637..346C}}.
With just a single Lofar station \citep[actually, the Kilpisj\"{a}rvi Atmospheric Imaging Receiver Array (KAIRA)][which uses Lofar technology]{2015ITGRS..53.1440M} it is possible to infer both the height and velocity of the scattering screen.

\citet{2020JSWSC..10...10F} demonstrate both that sufficiently active ionospheres occur at mid-latitudes, and also that multiple scattering screens can be resolved via this technique.
Furthermore, there is added value in also observing with multiple stations spread over several kilometres.

\subsection{``Multi-station'' amplitude scintillation}
\label{subsec:multi_station}
Since amplitude scintillation can be measured with a single Lofar or SKA-low station, and baselines approaching or exceeding $r_F$ provide semi-independent information on the diffraction pattern, utilising multiple stations can map out the diffraction pattern in detail.

As noted in the introduction there is commonality between observing trans-ionospheric signals in this way, and observing lower-frequency signals reflected off the bottomside of the ionosphere.
A similar approach is commonly used for interplanetary scintillation (IPS) studies (\citealp{1967Natur.213..343D}; see also discussion in \citealp{Chhetri01.2026.SKA} in this volume and references therein), and even interstellar scintillation \citep{2000aprs.conf..147J}.
For IPS, this technique is widely used, since the scintillation pattern observed by different locations on the ground, with Heliosphere piercepoints at different distances from the Sun, see strongly correlated scintillation patterns, but with a delay induced by the travel time of the solar wind from one piercepoint to the other.
The cross-correlation therefore provides a direct measurement of the motion of the diffraction pattern on the ground, which in turn provides a robust velocity measurement of the solar wind.
This is also the case for ionospheric studies; however, as first noted by \citet{1955PPSB...68..493S}, caution must be taken due to the extreme anisotropy of the scattering, and therefore of the diffraction pattern.
With small numbers of stations this can easily lead to erroneous speeds.

This is not a problem for modern interferometers.
On the contrary, \citet[][\fig~14]{10.1051/swsc/2025052} demonstrate how, with its large number of baselines, Lofar can be used to measure this field-aligned anisotropy \emph{and} the velocity of the diffraction pattern on the ground.
\citep[see also][]{10.22541/essoar.176169469.93512147/v1} 

A further consideration for velocity measurements with astrophysical sources is the motion of the piercepoint as the Earth rotates.
This is a useful feature, since it ensures that the diffraction pattern sweeps over a ground-based instrument, ensuring that even structures that are static (with respect to magnetic field lines and to the ground) still cause scintillation.
The piercepoint velocity, its height dependence, and how scintillation velocity measurements can be compared with velocities determined via other instruments is discussed in detail by \citet{2014JGRA..11910544F}.

The wide use of this approach with IPS for many decades underlines the wealth of information that can be gleaned from discrete spatial measurements of a time-varying 2D diffraction pattern, even with a small number of baselines.
\citet{1996Natur.379..429G} showed that different solar wind streams, with different velocities, could be resolved along the line of sight with a sufficiently long baseline, and this work continues with the much larger number of baselines provided by International Lofar \citep{2023AdSpR..72.5311F}.
\citet{1971A&A....10..306L} introduced a model that gives each Fourier component of the diffraction pattern its own velocity, with spread of these velocities characterised by a ``random component''.
\citet{2005JGRA..110.3101H} have demonstrated that this model can be use to probe the detailed physics of the microturbulence of the solar wind near the Sun.
Similar techniques should be applicable to the ionosphere.

\subsection{``Visibility scintillation''}
\label{subsec:visibility_scint}
In the previous section, each element of the ``multi-station'' array is used to provide a separate amplitude timeseries for the observed source(s).
In the case of Lofar and most other astronomical radio instruments, this is a repurposing of a set of elements that are also capable (indeed designed) to function together as an interferometer.
This means that multi-station amplitude scintillation measurements make no use of the interferometric phase.
\citet{1972ApJ...174..181C} provide a theoretical framework for working instead with interferometric visibilities, which, to our knowledge has never been tested.

It should be noted that the Cronyn approach implicitly envisions observing a single source.
Visibilities are a linear sum of all individual sources within the Field of View (FoV).
The visibilities of individual sources, at least the brighter ones, might be measured by iterative subtraction from the visibilities \citep{2008ISTSP...2..707M}.
Individual sources can also be isolated in the visibility domain via time- and frequency averaging \citep{2018MNRAS.478.2337R}, although care needs to be taken not to average over the frequency and temporal structure of the diffraction pattern itself.
An alternative approach would be to Fourier transform a ``cut-out'' of the image back to the visibility domain.

We note also that the secondary spectrum approach described in \sect~\ref{subsubsec:2spectrum} has also been generalised to an interferometric approach by \citet{2010ApJ...708..232B}; demonstrating, at least for the case of strong scattering of pulsars by the interstellar medium, that the amplitude and interferometric phase provide complementary information about the scattering screen.

\section{Connecting observations to simulations}
\label{sec:simulations}
Many simulations of the ionosphere have been made which seek to measure the impact of the ionosphere on radio interferometers.
We note that these tend to assume phase effects only \citep{2021PASA...38...28C,2025MNRAS.543.1092B},
and with a few exceptions \citep{2010ApJ...718..963K,2018A&A...615A.179D}, discussion of ionspheric amplitude scintillation is limited to ionospheric studies, although the impact of amplitude scintillation on transient searches has been noted \citep{2025PASA...42..129H}.

However, accurate calculation of phase and amplitude scintillation has been computationally feasible for some time.
For instance, the package \texttt{scintools} \citep{2010ApJ...717.1206C,2020ApJ...904..104R} is widely used for studies of the interstellar medium, particularly pulsar scintillometry.
\begin{figure}[ht]
    \centering
	\includegraphics[width=1.0\textwidth]{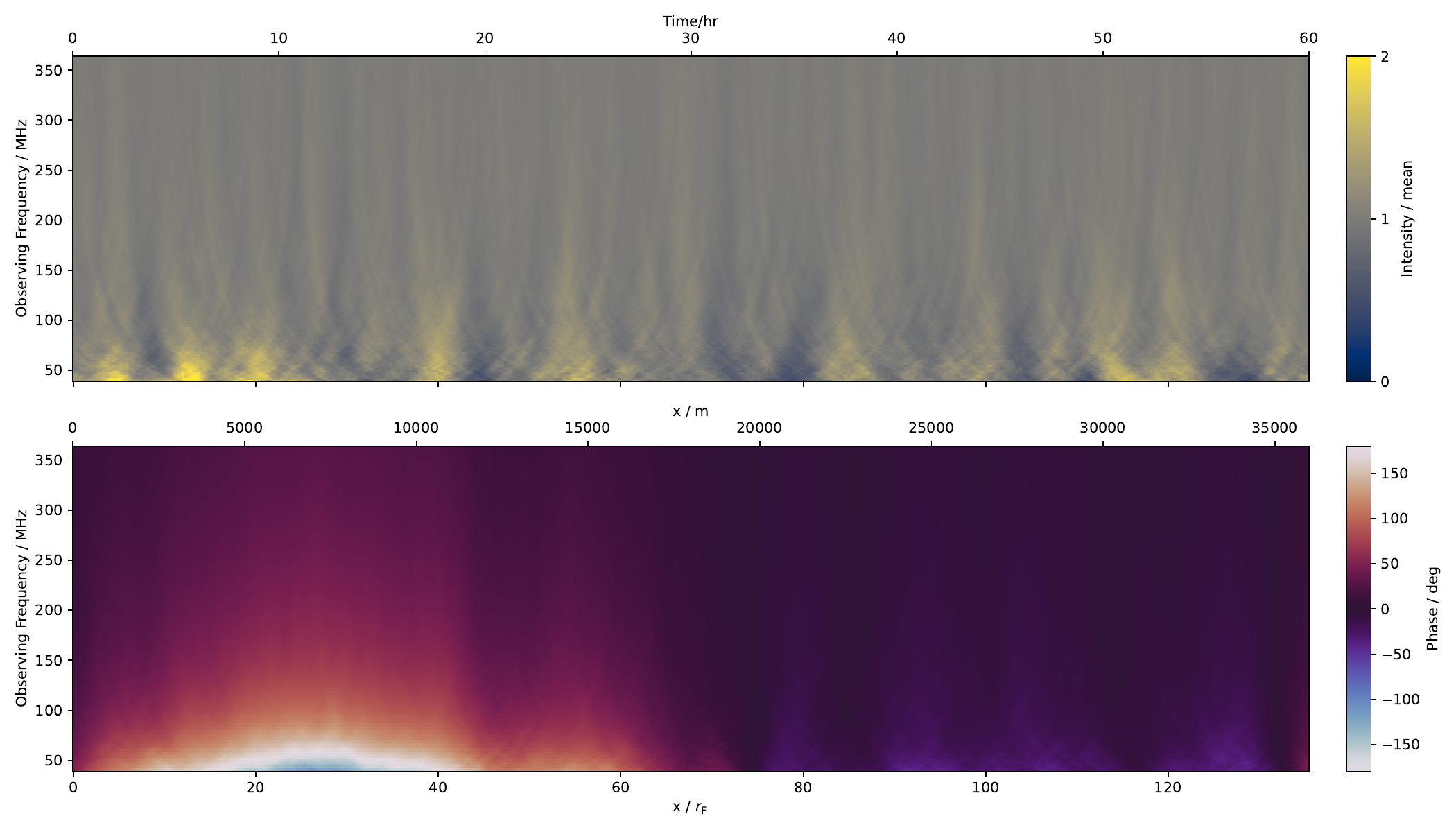} \\
	\includegraphics[width=1.0\columnwidth]{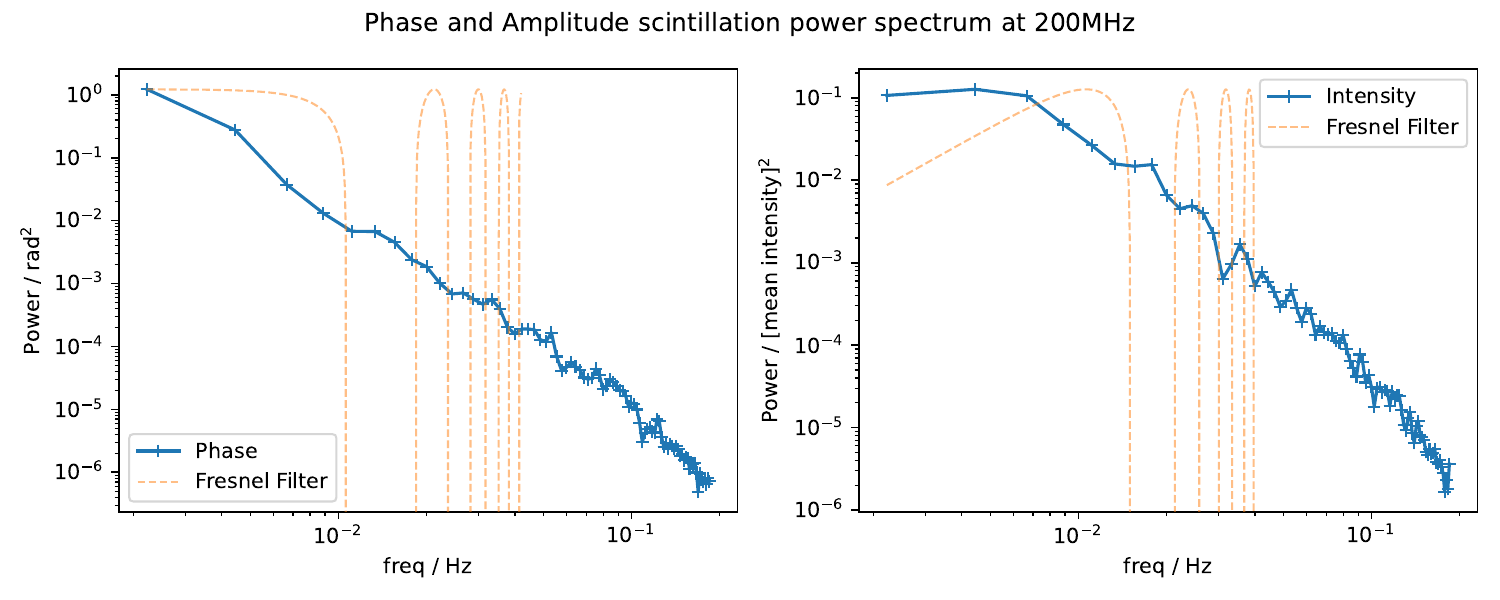}
	\caption{Output of \texttt{scintools} package amplitude and phase scintillation for a single ground location for slightly disturbed ionospheric conditions.
    Isotropic Kolmogorov turbulence with $r_\mathrm{diff}$=5\,km at 150\,MHz, ionospheric height of 300\,km is assumed.
    Top two panels: amplitude and phase respectively across the full SKA band.
    The three x-axis scales are common to both plots and are connected by an assumed piercepoint velocity of 20\,m\,s$^{-1}$, and $r_\mathrm{F}$=265\,m (that of the central frequency of 200\,MHz).
    Amplitude transfer function saturates at twice the mean intensity. Bottom two panels: Phase and amplitude power spectra for 200\,MHz.}
    \label{fig:scintools_spec}
\end{figure}

In order to show that it is useful for ionospheric studies also, we have used \texttt{scintools} to produce a one-hour dynamic spectrum across the full SKA band.
This is presented in \fig~\ref{fig:scintools_spec}, and simulation parameters are given in the caption.
The chosen scintillation strength corresponds to the 90th percentile of ionospheric activity observed by \citet{2016RaSc...51..927M}.

Note that the absolute phase is not something that can typically be measured by an interferometer \citep[unless the signal being detected is phase-stable; see e.g.][]{2014A&A...564A...4M}.
However, the length scale (in metres) gives a picture of how the phase changes across a particular baseline, or 1-dimensionally across an entire array; although multiple simulations might be needed to give the full picture for anisotropic turbulence.

Nonetheless it captures important behaviours.
Particularly clear is the striking effect of the Fresnel filter on amplitude scintillation\footnote{Discussion of the Fresnel filter, as well as the effect of finite screen depth can be found in \citet{wheelon2003} ~\Figs~3.22, 7.1 and 7.2 and accompanying text. This difference between phase and amplitude scintillation can also be understood if we consider the random walk in the complex plane of the phase vector. Assuming that the source is intrinsically stable in intensity, the amplitude is constrained by conservation of energy to vary about the mean value; however, the phase has no such constraint.}, namely the insensitivity of amplitude scintillation to larger-scale structures.
This is shown in the intensity power spectrum, where the turnover at low frequencies is clearly seen (although not dramatic in log-log space) but is even more obvious in the dynamic spectrum.

The simulation runs in seconds on a laptop, and so it is very easy to explore the effects of anisotropy, scattering strength, turbulence power law index, etc. on the scintillation of a single source at a single point on the ground.

A single, thin, frozen screen is not sufficient to capture all behaviours, and a state-of-the art framework allows the effects of the 3D ionospheric electron density and magnetic field to be propagated through to the amplitude and phase at the location of ground-based sensors \citep{2014JGRA..119.4026D}.
By generating a 3D realisation of ionospheric turbulence, forward propagating it to GNSS observables, and comparing with observations, \citet{2016JGRA..121.9188D} have shown that it is possible to iteratively solve for fundamental parameters such as thickness of the turbulent layer, and turbulence spectral index and axial ratio.

Furthermore, \citet{2019GeoRL..46.4564D} have coupled the electromagnetic wave propagation model to a physics-based plasma model \citep{2012JGRA..117.6306Z}. This requires a volume $\sim$100\,km on a side to be simulated with a spatial resolution $\sim$200\,m and time resolution of the simulations of 0.5\,s.
However, the output is the complex received signal across a 2D ground plane $\sim$10\,km on a side.
The received visibility for each baseline within this 2D ground plane can then be easily computed, meaning that with minimal effort, this system of models can be made to produce interferometric observables.

\section{Ionospheric science with SKA-Low}
\label{sec:roadmap}

\subsection{Applying ionosphere observation techniques to SKA-Low}
\begin{figure}[ht]
    \centering
	\includegraphics[width=1.0\columnwidth]{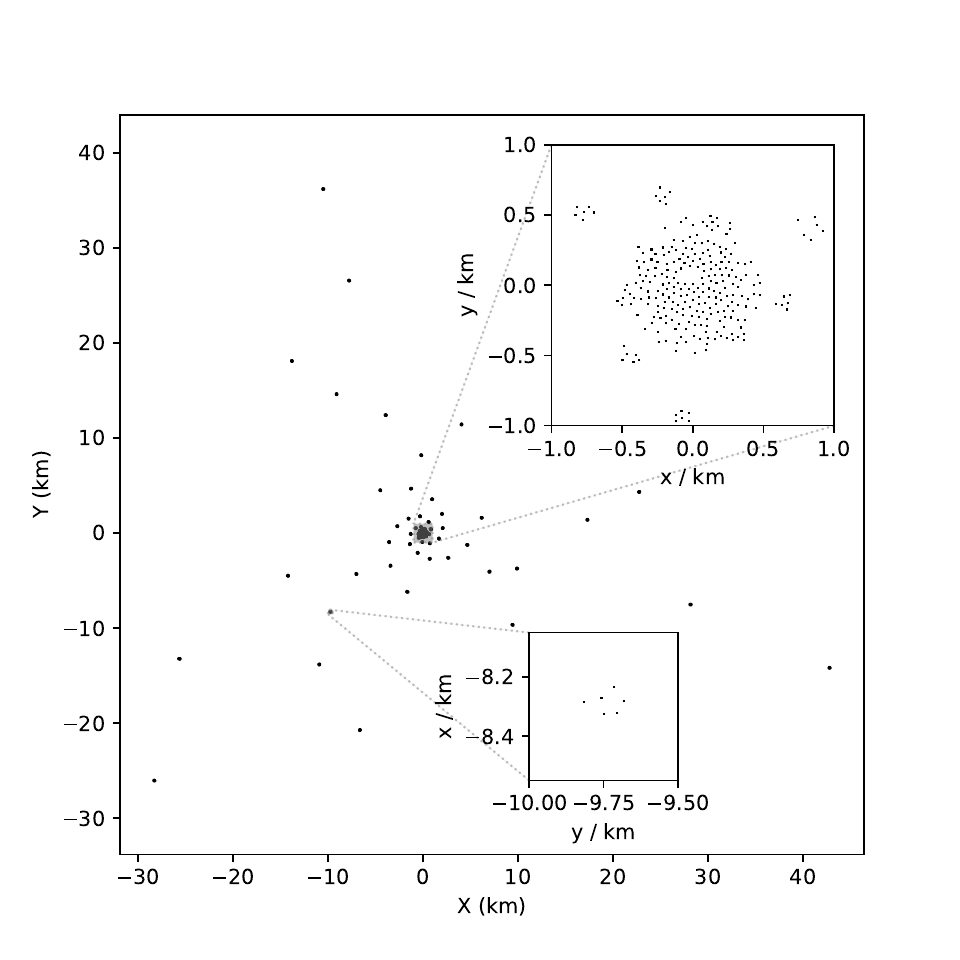}
	\caption{Layout of the SKA-low (AA4 configuration) with inserts showing the high-density core (top right), and the stations of a single spiral arm cluster (bottom right.)}
    \label{fig:layout}
\end{figure}
The SKA-low AA4 will consist of 512 individual stations, each consisting in turn of 256 antennas covering a circular patch approx. 40\,m in diameter.
The layout of the 512 stations is shown in \fig~\ref{fig:layout}.
Approximately half of the stations are in the core, within a 500\,m radius of the array centre.
The remainder are arranged into clusters of 6 stations, which extend to a radius of 40\,km.
Each cluster is contained within approximately 100\,m diameter.

According to the SKA sensitivity calculator, for an extragalactic field, a single SKA-low baseline can achieve a sensitivity of $\sim$133\,mJy with 30\,MHz bandwidth and 4\,s integration time (the sensitivity of a single station as an interferometer should be very similar).
Assuming an observation lasting 100 scintillation timescales (approximately 800\,s), this gives us a signal to noise (S/N) of 4 for a 4\,Jy source, which would be sufficient to measure a scintillation index.
A S/N of 20, which is sufficient for analysis of the temporal power spectrum \citep{2019SpWea..17.1114C} is attainable for a 20\,Jy source.
Depending on frequency, sources of this brightness occur at the rate of one per resolution unit of a single SKA station, so there will be many sources that can be analysed; however the precise number will be strongly frequency dependent, a number of factors could impact (classical and sidelobe confusion) and detailed empirical studies will be required in order to determine how many sources can be used in single station and multi-station analysis, and whether refractive shifts can be determined from a single outer station cluster.

\subsection{Summary of observing requirements}
\citet{1972ApJ...174..181C} gives an overview of the requirements for observing visibility scintillation (which they note become amplitude scintillations in the zero-baseline limit):
The instrument must be otherwise phase stable (fluctuations in phase after correcting for geometric effects (e.g. Earth rotation) must be attributable to the ionosphere); baseline lengths should be of order of, or longer than the spatial scale of the diffraction pattern on the ground, integration time should be substantially less than the timescale over which the diffraction pattern changes at a given location. Finally, for statistical stability it is important to observe over $\sim$100 scintillation timescales.
All of these requirements should be possible to fulfil with the SKA-low.

We would add that it is important that the collecting area of each element is also sufficiently small that phase and amplitude changes across it are small. 
This is likely to be true in almost all cases for the SKA-low.
\subsubsection{Data products required}
Methods outlined in \sects~\ref{subsec:shift} would utilise fairly standard image products, possibly with sub-arrays for parallax.
Methods in \sects~\ref{subsec:baseline} and \ref{subsec:visibility_scint} would require access to visibilities, though it is possible that Fourier inversion of dirty images may suffice, experimentation would be required.

For measuring amplitude scintillation (\sect~\ref{subsec:scint}), we envision using all-sky images formed from using individual stations in an interferometry mode as all-sky monitors.
Imaging cadence of $\sim$5\,s would probably be sufficient, a single channel with fractional bandwidth of $<10$\% would provide useful results, though there would certainly be value in having multiple spectral channels, especially at the bottom of the band, and under extreme ionospheric conditions, since dynamic spectra of strong scintillation provide more information than a single timeseries (and allow the use of the secondary spectrum approach).

A single station would provide interesting results, but much more interesting would be the possibility of using multiple stations in the core in order to use the multi-station approach outlined in \sect~\ref{subsec:multi_station}.

\subsubsection{Stations required}
At least initially, stations in the core, or a subset of them would provide a wealth of data.
$\sim$ 1\,km baselines would be sufficent to measure refractive shifts, though slightly longer baselines would be advantageous, particularly if parallactic imaging is to be employed.

There are also many cases where clusters in the spiral arms can provide useful information.
These will be required in order to use baseline-based measurements in order to construct the 2D ionospheric structure function beyond 1\,km.
Additionally, the $\sim$5$^\circ$ field of view of the SKA corresponds to a 25\,km linear size on the ionosphere at 300\,km. 
By observing sources from the core and the spiral arms, it is possible to pierce this volume of the ionosphere with lines of sight at multiple angles, potentially allowing a full 3D reconstruction.

With regard to more or fewer stations relative to the AA4 design, we first note that ionospheric measurements are possible with any subset of the array, and detailed ionospheric measurements could be made throughout the commissioning period, with useful measurements being possible even with a single station.
Ionospheric measurements are unlikely to be affected by a modest reduction of stations in the core; however the use of an outer cluster stations to do refractive ionospheric measurements is probably already marginal, so further reductions would be problematic.

The slightly smaller number of core stations in AA* is unlikely to make much difference for ionospheric science, though the increased number of outer stations in AA4 relative to AA* will be very useful for refractive shift measurements, particularly for parallax measurements.
While longer baselines and greater density of stations outside the core would be welcome for characterising the ionosphere on the larger scales in greater detail, the AA4 layout is probably sufficient.
For developments beyond AA4, ionospheric science would be better served by putting resources into flexibility in science data processing and products: capabilities for station-level imaging for the widest bandwidths possible for larger numbers of stations, and larger numbers of station beams to enable a wider field of view, at least for a subset of stations.
Also, as already noted, provision of visibilities to the observer would permit a wider range of analyses than if we are forced to work with image data alone.

\section{Summary and Conclusions}
In this chapter we have tried to summarise the broad sweep of ionospheric studies that have been possible with the low-frequency precursors to the SKA (\sect~\ref{sec:methods}).
There is a great deal more work, and potential techniques that remain to be explored even with these precursor instruments.
Of particular interest is the potential for connecting interferometer observables to sophisticated 3D simulations of the ionosphere (\sect~\ref{sec:simulations}).
As outlined in \sect~\ref{sec:roadmap}, the SKA-low layout offers enormous opportunities for applying these approaches and more to explore the mid-latitude ionosphere in unprecedented detail. 

\bibliographystyle{abbrvnat-maxbibnames4}
\bibliography{chapter}

\end{document}